\documentclass[12pt, prd, showpacs]{revtex4}
\usepackage{amssymb}

\input{tcilatex}

\begin{document}

\title{N-spheres in general relativity: regular black holes without apparent
horizons, static wormholes with event horizons and gravastars with a
tube-like core}
\author{O. B. Zaslavskii}
\affiliation{Department of Mechanics and Mathematics, Kharkov V.N.Karazin National
University, \\
Svoboda Square 4, Kharkov 61077, Ukraine}
\email{ozaslav@kharkov.ua}

\begin{abstract}
We consider a way to avoid black hole singularities by gluing a black hole
exterior to an interior with a tube-like\ geometry consisting of a direct
product of two-dimensional AdS, dS, or Rindler spacetime with a two-sphere
of constant radius. As a result we obtain a spacetime with either
"cosmological" or "acceleration" (event) horizons but without an apparent
horizon. The inner region is everywhere regular and supported by matter with
the vacuum-like equation of state $p_{r}+\rho =0$ where $p_{r}=T_{r}^{r}$ is
the longitudinal pressure, $\rho =-T_{0}^{0}$ is the energy density, $T_{\mu
}^{\nu }$ is the stress-energy tensor. When the surface of gluing approaches
the horizon, surface stresses vanish, while $p_{r}$ may acquire a finite
jump on the boundary. Such composite spacetimes accumulate an infinitely
large amount of matter inside the horizon but reveal themselves for an
external observer as a sphere of a finite ADM mass and size. If the throat
of the inner region is glued to two black hole exteriors, one obtains a
wormhole of an arbitrarily large length. Wormholes under discussion are
static but not traversable, so the null energy condition is not violated. In
particular, they include the case with an infinite proper distance to the
throat. We construct also gravastars with an infinite tube as a core and
traversable wormholes connected by a finite tube-like region.
\end{abstract}

\maketitle

% It is always \today, today, but any date may be explicitly specified


The nature of inner structure of black holes and the problem of their
singularity is one of central issues in black hole physics \cite{fn}.
Different attempts were undertaken to remove a singularity by making
composite spacetimes that reveal themselves as a black hole for an external
observer but contain a regular inner region. In doing so, the special role
is played by the de Sitter (dS) metric which is supposed to mimic
vacuum-like media \cite{gliner}. Here, different possibilities arise: one
can (1) replace the part of a black hole metric by the dS one inside the
horizon \cite{fmm}, (2a) consider some conceivable distribution of matter
that interpolates smoothly between the Schwarzschild and dS metrics \cite%
{dym} or (2b) sew two exact solutions smoothly due to special fine-tuning of
parameters, the composite spacetime having a horizon \cite{em}, (3) sew the
black hole and dS metrics (or its generalization) outside the horizon in
such a way that the horizon does not form at all (so-called gravastars \cite%
{maz}).

All the aforementioned approaches assume that the central singularity is
replaced by some regular interior in which this singularity is smoothed out
in the centre. In the present work we suggest a quite different way - to get
rid off the singularity in the centre by simply getting rid of the centre by
itself. This idea is realized by sewing an outer black hole region with
spacetimes having no centre of symmetry such as Bertotti-Robinson (BR) \cite%
{br} or Nariai metric \cite{nar} or or the direct geometrical product of
two-dimensional Rindler spacetime and a fixed two-sphere (Rindler$_{2}$xS$%
_{2})$. For all such spacetime the algebraic structure of the stress energy
tensor $T_{0}^{0}=T_{r}^{r}$ is invariant under radial boosts similarly to
properties of metrics considered in \cite{dym}, \cite{em}. However,
spacetime structure is qualitatively different. In particular, the event
horizon is not accompanied by the apparent horizon. Apart from this, in some
particular examples the role of black hole horizons is played by the
acceleration ones that usually represent a pure kinematical effect and
disappear after passing to the proper chosen frame. We will see that such
composite spacetimes automatically possess one more important features:
although on the boundary stresses persist, they asymptotically vanish in the
limit as the shell approaches the horizon.

As far as the spacetime structure of the inner region is concerned, the
aforementioned options (1) and (2) correspond to T-regions in the sense that 
$\left( \nabla r\right) ^{2}<0\ $where $r$ is the areal radius (we use the
terminology of Ref. \cite{zln}). In the case (3) the interior spacetime
represents R region for which $\left( \nabla r\right) ^{2}>0$. In this
sense, our case occupies the intermediate$\ $position since $\left( \nabla
r\right) ^{2}=0$ inside just because of constancy of $r$. For brevity, we
will call it N-region. Thus, the whole spacetime consists of gluing one R
and one N region.

Consider the static metric%
\begin{equation}
ds^{2}=-dt^{2}f+dl^{2}+r^{2}(l)(d\theta ^{2}+d\phi ^{2}\sin ^{2}\theta )%
\text{, }f=b^{2}  \label{ml}
\end{equation}%
If $r$ can be chosen as a variable (that is not always the case - see
below), it can be rewritten in the equivalent form%
\begin{equation}
ds^{2}=-dt^{2}f+\frac{dr^{2}}{V}+r^{2}(d\theta ^{2}+d\phi ^{2}\sin
^{2}\theta )\text{, }V=\left( \frac{dr}{dl}\right) ^{2}\text{.}  \label{mr}
\end{equation}

We would like to glue to different spacetimes along the time-like surface
(shell) $r=r_{0}$. Following the general formalism \cite{isr}, one can write 
\begin{equation}
8\pi S_{\mu }^{\nu }=[K_{\mu }^{\nu }]-\delta _{\mu }^{\nu }[K]\text{,}
\label{k}
\end{equation}%
where $S_{\mu }^{\nu }\equiv \int_{r_{0}-0}^{r_{0}+0}dl\tilde{T}_{\mu }^{\nu
}$, $\tilde{T}_{\mu }^{\nu }$ is the stress-energy tensor of the shell, $%
K_{\mu }^{\nu }$ is the tensor of the extrinsic curvature calculated on the
surface $r=r_{0}$, $K=K_{i}^{i}$ ($i=0,2,3$) and $[...]=(...)_{+}-(...)_{-}$%
, signs "+" and "-" correspond to the outer and inner regions, respectively.
If $[K_{\mu }^{\nu }]=0$, the quantity $S_{\mu }^{\nu }$ vanishes and both
regions match smoothly. Calculating $K_{\mu }^{\nu }$ from (\ref{k}) one can
easily obtain

\begin{equation}
K_{0}^{0}=-\frac{b^{\prime }}{b}\text{, }K_{2}^{2}=-\frac{r^{\prime }}{r}%
=K_{3}^{3}\text{, }K=-\frac{2r^{\prime }}{r}-\frac{b^{\prime }}{b}\text{,}
\end{equation}%
where prime denotes differentiation with respect to the proper length $l$.
We have ($\sigma \equiv -S_{0}^{0}$, $\Theta \equiv -S_{2}^{2})$%
\begin{equation}
8\pi \Theta =[K_{0}^{0}]+[K_{2}^{2}]=-\frac{(br)_{+}^{\prime
}-(br)_{-}^{\prime }}{br}\text{,}  \label{s22}
\end{equation}%
\begin{equation}
8\pi \sigma =2[K_{2}^{2}]=-\frac{2(r_{+}^{\prime }-r_{-}^{\prime })}{r}\text{%
.}  \label{s00}
\end{equation}

Let the stress-energy tensor be represented in the form $T_{\mu }^{\nu
}=diag(-\rho $, $p_{r}$, $p_{\perp }$, $p_{\perp })$. If $b^{\prime }$ has
different signs from both sides of the boundary (like it happens for
gravastars \cite{maz} or their simplified version \cite{visgr}), the tensor $%
S_{\mu }^{\nu }$ does not vanish and, moreover, as the boundary approaches
the horizon, the stresses grow unbound. If $b_{+}^{\prime }$ and $%
b_{-}^{\prime }$ have the same sign, one can combine known exact solutions
to obtain smooth gluing \cite{em}. We consider now a quite different
situation. We choose the metric of interior "-" to obey the Einstein
equations with $r=r_{0}=const.$ Then it follows from $00$ and $11$ equations
that $\rho ^{-}=-p_{r}^{-}=\frac{1}{8\pi r_{0}^{2}}$ and $22$ equation gives
us $\frac{b^{\prime \prime }}{b}=8\pi p_{\perp }^{-}$, where $(...)_{\pm
}\equiv \lim (...)_{r\rightarrow r_{0}\pm 0}$. Thus, the interior should be
vacuum-like in the sense that $\rho +p_{r}=0$, and there are three different
cases depending on the sign of $p_{\perp }$. If (1) $p_{\perp }>0$, then (a) 
$b=a\sinh \kappa l$, where $a$ is a constant, $\kappa ^{2}=8\pi p_{\perp }$,
(b) $b=a\exp (\kappa l)$ or (c) $b=a\cosh \kappa l$. If (2) $p_{\perp }<0$,
by a suitable linear transformation of $l$ we can achieve $b=a\sin \kappa l$
with $\kappa ^{2}=-8\pi p_{\perp }$, if (3) $p_{\perp }=0$, we have (a) $%
b=al $ or (b) $b=a$. Particular examples of corresponding physical sources
are electromagnetic field (case 1 with $p_{\perp }=\rho $ - BR solution),
cosmological constant (case 2 with $p_{\perp }=-\rho $ - Nariai solution),
string dust \cite{sdust} (case 3).

In all these cases formulas (\ref{s22}) and (\ref{s00}) can be rewritten as%
\begin{equation}
4\pi \sigma =-\frac{\sqrt{V_{+}}}{r_{0}}\text{,}
\end{equation}%
\begin{equation}
8\pi \Theta =-\frac{\sqrt{V_{+}}}{r_{0}}-[\frac{\partial \ln b}{\partial l}]%
\text{.}
\end{equation}

For any gluing outside the horizon one cannot glue smoothly the N-region
with the R-one (in agreement with the remark about Nariai solution in Sec.
IVa of \cite{em}) but, nonetheless, we will see now that in the horizon
limit both $\sigma $ and $\Theta $ asymptotically vanish. Let us discuss
separately the cases when the exterior represents (i) a non-extremal black
hole, (ii) an extremal one. Let $r_{0}\rightarrow r_{h}$, where $r_{h}$
corresponds to the horizon. Consider first the case (i). Then $b_{h}^{\prime
}\neq 0$ by definition and we have in the "+" region for small $l$ the
asymptotic expansion $b=b_{h}^{\prime }l[1+O(l^{2})]$. The quantity $\sqrt{%
V(r)}$ behaves like $\sqrt{r-r_{h}}\sim l$. We glue the "+" region with
versions 1a), 2) or 3a) of the "-" region. Then $\frac{\partial \ln b}{%
\partial l}$ has the same asymptotic form $\frac{\partial \ln b}{\partial l}=%
\frac{1}{l}+O(l)$ on both sides of the shell, in the "+" region $\frac{%
\partial r}{\partial l}\rightarrow 0$ and in the "-" region $\frac{\partial r%
}{\partial l}=0$ exactly. As a result, we obtain that $\sigma $, $\Theta
\sim l\rightarrow 0$. It is worth stressing that one can glue any two
spacetimes of the kind under discussion.

Consider case 3a) as an example. Inside the shell, one can introduce the new
variables according to $X=l\cosh (a\tau )$, $T=l\sinh a\tau $, perform the
transformation and obtain in the interior the new metric of the same form
but with the new $b=const$, in other words the Minkowski two-dimensional
spacetime in agreement with well known relation between the Rindler and
Minkowski spacetimes, the total metric being $%
ds^{2}=-dT^{2}+dX^{2}+r_{0}^{2}(d\theta ^{2}+d\phi ^{2}\sin ^{2}\theta )$.
Thus, the two-dimensional part mimics the empty space but because of the
angular part the four-dimensional spacetime is curved. As is well known, in
the two-dimensional flat spacetime the family of Rindler observers following
the trajectories $l=const$ covers not the whole spacetime but only one
quadrant bounded by past and future acceleration horizons. If an observer
passes to $X$, $T$ frame, the acceleration horizon in accordance with its
pure kinematic nature disappears and this new frame covers all $X$ - $T$
manifold, so that all signals can escape to corresponding infinity. (In
cases 1a and 1b we are faced with the AdS two-dimensional geometry that also
possesses acceleration horizons, in case 2 the geometry of two-dimensional
part is of the dS type and the horizon is "cosmological".)

However, now the four-dimensional nature of spacetimes comes into play.
Usually, the Carter-Penrose diagrams representing the structure of spacetime
are pure two-dimensional, with the reservation that each point represents a
two sphere of the areal radius $r$. In doing so, the coordinate $r$ plays
the double role: it enters spacetime diagrams and it measures the surface
area. In particular, for the case of asymptotically flat spacetimes (that
embraces Schwarzschild and Reissner-Norstr\"{o}m black holes) $r\rightarrow
\infty $ at spatial and null infinity. Meanwhile, for the case under
discussion coordinates $X$, $T$ (or similar coordinates for the BR metric)
have nothing in common with coordinates $r$, $t$ of asymptotically flat
spacetime since $r=r_{0}=const$ inside the N-region. Therefore, although
inside the shell only an acceleration horizon is present, signals from the
interior cannot reach an observer at infinity (and even an observer with a
finite $l$ between the horizon and the shell). As a result, we have a black
hole in the sense that there is a spacetime region from which light cannot
escape to infinity. As the quantity $r$ is constant inside, there are no
trapped surfaces at all. Thus, we obatain a black hole with an event horizon
but without apparent horizons.

Up to now, we considered the non-extremal horizons. In the extremal case
(ii) we must select the only suitable candidate for smooth gluing, case 1b)
with $b_{-}=a\exp (\kappa l)$, so that $\frac{\partial \ln b_{-}}{\partial l}%
=\kappa $. Let in the outer region the metric have the asymtotics typical of
extremal black holes: $b_{+}=B(r-r_{h})+O(r-r_{h})^{2}$, $%
V=A^{-2}(r-r_{h})^{2}+O(r-r_{h})^{3}$, where $A$ and $B$ are constants. Then 
$b_{+}\sim (r-r_{h})\sim \exp (\frac{l}{A})[1+O(\exp (\frac{l}{A}))]$, $%
l\rightarrow -\infty $ and $\frac{\partial \ln b_{+}}{\partial l}=\frac{1}{A}%
+O[\exp (\frac{l}{A})]$. In the same manner, one can easily calculate
stresses and obtain that they are proportional to $\exp (\frac{l}{A})$ and
vanish in the limit under consideration, provided $\kappa =\frac{1}{A}$,
whence $p_{\perp }^{-}=\frac{1}{8\pi A^{2}}$. Consider, for simplicity, the
BR spacetime. Then $A=r_{h}$, $b=\sinh \frac{l}{r_{h}}$. By boosts in the
radial direction satisfying 
\begin{equation}
\sinh y=\frac{1}{2\xi }(t^{2}-\xi ^{2}+1)\text{, }\cos T\cosh y=\frac{1}{%
2\xi }(t^{2}-\xi ^{2}-1)\text{, }\xi \equiv e^{-l}\text{,}
\end{equation}%
where for a moment we put for simplicity $r_{h}=1$, we may achieve $b$ to
have the form 1c) according to known properties of BR spacetime, so that $%
ds^{2}=-dT^{2}\cosh ^{2}y+dy^{2}+d\theta ^{2}+\sin ^{2}\theta d\phi ^{2}$.
It follows from (\ref{mr}) and 00 Einstein equation that the Hawking
temperature $T_{H}=\frac{1}{4\pi r_{+}}(1-\frac{\rho ^{+}}{\rho ^{-}})\exp
(\psi _{+})$, where $\psi _{+}=4\pi \int_{\infty }^{r_{+}}dr\frac{%
r(T_{r}^{r}-T_{0}^{0})}{V}.$ If the horizon is extremal, $T_{H}=0$ and $\rho
^{+}=\rho ^{-}=\frac{1}{8\pi r_{h}^{2}}$. With the regularity conditions on
the horizon $T_{r}^{r}-T_{0}^{0}=0$, one obtains that $p_{r}^{+}=-\rho
^{+}=-\rho ^{-}=p_{r}^{-}$. Thus, the radial pressure is continuous. (In the
particular case when both inside and outside $p_{\perp }=\rho =$ $-$ $p_{r}=%
\frac{e^{2}}{8\pi r^{4}}$ where $e$ is an electric charge, the results of 
\cite{vf} for sewing the BR spacetime with the extremal Reissner-Norstr\"{o}%
m metric are reproduced.) However, it does not necessarily hold for
non-extremal horizons in which case radial pressure can acquire a jump. For
example, this happens in the case of the outer Schwarzschild metric: in the
"+" region $p_{r}^{+}=0$ but in the "-" region $p_{r}^{-}=-\rho ^{-}\neq 0$.
Thus, the tangential stresses asymptotically vanish but the jump in $p_{r}$
does not. Such a seemingly paradoxical combination is easily explained if
one invokes the conservation law $T_{\mu ;\eta }^{\nu }=0$ with $\mu =l$,
whence $(\sqrt{-g}T_{1}^{1})^{\prime }=r^{2}b^{\prime }T_{0}^{0}+2r^{\prime
}brT_{2}^{2}$. Then it is clear that it is combination $\sqrt{-g}T_{1}^{1}$
which enters the expression for the jump due to jumps in $b^{\prime }$ and $%
r^{\prime }$. Usually, $\sqrt{-g}\neq 0$ and, if other components are
continuous across the shell, continuity of $\sqrt{-g}T_{1}^{1}$ is
equivalent to the continuity of $T_{1}^{1}$. However, as the shell
approaches the horizon, $\sqrt{-g}\sim b\sim l\rightarrow 0$. Therefore, the
jump in $p_{r}$ is compatible with the continuity of $\sqrt{-g}T_{1}^{1}$.

The composite spacetimes under discussion have one more interesting property
connected with the gravitational mass defect. The gravitational mass
measured in the outer region is equal to $m(r)=m(r_{0})+4\pi
\int_{r_{0}}^{r}dr\rho r^{2}$, the ADM mass $m(\infty )$ being finite since
outside the shell matter is supposed to be bounded within some compact
region or the density $\rho $ decreases rapidly enough. In the limit $r_{0}$ 
$\rightarrow r_{h}$ the mass $m(r_{0})$ tends to $m(r_{h})$ and is finite.
However, the total proper mass $m_{p}=4\pi \int dl\rho ^{2}$ measured on the
hypersurface $T=const$ in the tube under the shell at $r_{0}$, obviously,
diverges. It is not surprising that $m_{p}$ is infinite for an extremal
horizon in the outer region since $l$ diverges ($dl\sim \frac{dr}{r-r_{+}}$%
). Meanwhile, now $m_{p}$ diverges also for the \textit{non-extremal }%
horizons due to an infinite tube inside the shell (for instance, as a result
of integration over $X$ in the two-dimensional Minkowski case). To some
extent, it resembles the so-called T-spheres that can reveal themselves as a
body of a finite mass and size for an external observer whereas they bind an
infinite amount of matter inside the horizon \cite{rub} (see also \cite{nov2}%
). By analogy, we call such objects N-spheres. As there is no singular
centre here, N-spheres can be considered as realization of Wheeler's idea of
"mass without mass", alternative to T-spheres \cite{rub}. However, we would
like to stress that, while in the case of T-region matter collapses or
starts from the singular state, in our case the interior is perfectly
regular. As the media with the equation of state $p+\rho =0$ can be thought
of as gravitational vacuum condensate \cite{gliner}, \cite{dym}, \cite{maz},
the fact that an object with infinite "bare" energy reveals itself in
physical observations as a body with a finite energy, can be viewed as a
classical analogy of known properties of vacuum in quantum field theory.

Up to now we discussed gluing between two regions only. One can proceed
further and glue in the same manner another Schwarzschild (or extremal black
hole) region from the left, but again with the shell in the R-region.
Actually, we have some generalization of notion of wormholes \cite{mt}, \cite%
{vis} - with a throat of an arbitrary length lying in the N-region and
connecting two R-regions. Inside the throat the equation of state is exactly
vacuum-like $p_{r}+\rho =0$, the proper mass bounded inside the throat can
be made as large as one wishes (the configuration considered in \cite{phz}
tends to such a throat in some particular limit). The possibility of
extended throats ("hyperspatial tubes") for generic static wormholes was
briefly mentioned in \cite{visgen}, \cite{dav47}. We would like to stress
that in our case such objects are combined with the event horizons. It looks
natural to call wormholes with tube-like geometries inside "N-wormholes".

As a matter of fact, we have an object that interpolates between "ordinary"
black holes and wormholes. In particular, this reveals itself in the
following: the typical feature of black holes is the trapped surfaces while
the typical feature of wormholes is the "antitrapped" surfaces (see remarks
due to D. Page on p. 405 of Ref. \cite{mt} and \cite{antitr}). But now we
have neither first nor second case since $r=const$ in both directions inside
the horizon. N-wormholes under discussion are simultaneously static, not
traversable but safe for one-way travel (no tidal forces or spacetime
singularities occurs inside the tube). In particular, this includes the
extremal case, when the proper distance to throat is infinite. In the
absence of horizons, the spacetime would be geodesically complete, time of
travel would be infinite and there would be no wormhole at all. Now, thanks
to the horizon, the time is finite, so that an observer is able to go
through the tube but is unable to return.

The constructions under discussion contain horizons as a result of the
limiting procedure. In doing so, we used cases 1a), 1b), 2), 3a) for gluing.
Meanwhile, there exists alternative to it. Let us take, as an exterior, a
region from some traversable wormhole instead of a black hole and glue it to
the N-region. Then the horizon is present neither in the original spacetime
nor in the composite spacetime. To accomplish this, we should use cases 1c)
or 3b) complementary to our previous choice since for them $b\neq 0$ on the
throat and there are no horizons, as requested. Thus, in sum we exhaust all
possible cases 1) - 3). The composite spacetime in the case under current
discussion realizes literally the gravastar construction since there is no
horizon. It is natural to call it "N-gravastar" since it contains a core
with a tube inside. In contrast to original constructions \cite{maz}, where
surface stresses grow unbound as one approaches a would-be horizon, now
these stresses are not only finite but vanish at all. To see this, it is
worth noting that in the outer region $b^{\prime }=\frac{\partial b}{%
\partial r}r^{\prime }=0$ on the throat due to the factor $r$. It follows
from the explicit form of $b$ inside that in cases 1c) and 3b) $b^{\prime }=0
$ also in the N-region. In a similar way, $r^{\prime }=0$ both outside on
the throat and everywhere inside. As a result, stresses (\ref{s22}), (\ref%
{s00}) vanish. One can take the position of the would-be horizon as close to
the throat as one likes (for instance, one may take the metric similar to
that in eq.7 of \cite{dav47} with $b^{2}\sim (r-r_{0})^{2}+\varepsilon ^{2}$%
, $\varepsilon \rightarrow 0$) but this does not affect this circumstance.
Proceeding further in the same manner as before, one can accomplish gluing
from both sides of the N-region to obtain N-wormhole without horizons. By
construction, this kind of N-wormholes is traversable. Actually, it is
obtained with the help of cut- and paste technique like in Chapter 15 of
Ref. \cite{vis}. The difference consists in that now, instead of a thin
shell, we work with tubes of a finite (but arbitrarily large) length.

As is well known, the existence of static traversable wormholes entails, as
the necessary condition, the violation of NEC (null energy conditions) \cite%
{mt}, \cite{vis}. Meanwhile, in our case, this condition is satisfied
(although, on the verge) inside the N-region, \thinspace $p_{r}^{-}+\rho
^{-}=0$. If we consider traversable N-wormholes obtained by the surgery
based on cases 1c) or 3b), NEC\ is inevitably violated on the throat, $%
p_{r}^{+}+\rho ^{+}<0$ \cite{mt}, \cite{vis}. In the absence of horizons,
smooth gluing entails $p_{r}^{-}=p_{r}^{+}$, so that we obtain, as
by-product, that $\rho ^{+}<\rho ^{-}$. However, if we glue according to
prescriptions 1a), 1b), 2), 3a) (when the horizon is present), NEC is
marginally satisfied not only in the "-" region but also from the "+" side
of the throat. The difference can be understood as follows. One can easily
obtain from 00 and 11 Einstein equations that $G_{1}^{1}-G_{0}^{0}=\frac{%
2r^{\prime }b^{\prime }}{rb}-\frac{2r^{\prime \prime }}{r}$. If there is no
horizon, $b\neq 0$ and the first term vanishes on the throat due to the
factor $r^{\prime }$ (moreover, $b^{\prime }=\frac{\partial b}{\partial r}%
r^{\prime }$, so that $b^{\prime }$ also vanishes). As the throat is
supposed to be a minimum of $r$, the second derivative $r^{\prime \prime }>0$%
, so that $G_{1}^{1}-G_{0}^{0}=8\pi (p_{r}+\rho )<0$, and NEC is violated.
However, if $b\sim l$, $r-r_{h}\sim l^{2}$ with $l\rightarrow 0$
(non-extremal case) or $r-r_{h}\sim \exp (\frac{l}{r_{h}})$, $b\sim \exp (%
\frac{l}{r_{h}})$ with $l\rightarrow -\infty $ (extremal case), it follows
from the above expression that $p_{r}^{+}+\rho ^{+}\rightarrow 0$. Thus, NEC
in the "+" region is satisfied just due to the properties of the horizon. As
now a wormhole is not traversable, there is nothing wrong in that NEC is not
violated.

To summarize, we constructed composite objects that interpolate between
black holes and gravastars in that there is no horizon in the particular
solution obtained by gluing different regions of spacetime but the horizon
appears as a result of the limiting procedure when the object turns into
what we called a N-sphere. In doing so, we obtained event horizons without
apparent ones. Alternatively, we also obtained a gravastar with an infinite
tube as a core (N-gravastar). Generalization of the procedure under
consideration gave rise to objects interpolating between black holes and
wormholes (not traversable N-wormholes) or connecting two external regions
without horizons (traversable N-wormholes).

The type of geometry inside the N-region can be written as N$_{2}$xS$_{2}$
where N$_{2}$ is two-dimensional subspace - Rindler (if $p_{\perp }=0$), AdS
($p_{\perp }>0$) or dS ($p_{\perp }<0$) one. Correspondingly, there are
three possible types of N-spheres. As the geometry of the kind N$_{2}$xS$%
_{2} $ does not change its type when influenced by quantum backreaction (see 
\cite{back} and references [6], [8] - [15] therein), the whole construction
survives at the semi-classical level.

The constructions considered in the present paper can be also relevant for
higher-dimensional generalization of BR-like solutions \cite{vit}. In
particular, it concerns the issue of compactification of extra-dimensions in
Kaluza-Klein theories where flux tubes of constant cross-sections can arise
as compactified BR-like phase inside an uncompactified one \cite{gl}, \cite%
{flux}.


\begin{thebibliography}{99}
\bibitem{fn} V. P. Frolov and I. D. Novikov, \textit{Black hole Physics.
Basic Concepts and New Developments.} (Kluwer Academic Publishers,
Dordrecht/ Boston/ London, 1998).

\bibitem{gliner} E. B. Gliner, Sov. Phys. JETP \textbf{22}, 378 (1966).

\bibitem{fmm} V. P. Frolov, M. A.\ Markov and V. F. Mukhanov, Phys. Rev. D%
\textbf{\ 41} (1990) 383.

\bibitem{dym} I. Dymnikova, Gen. Rel. Grav. \textbf{24}, 235 (1992); Phys.\
Lett. B \textbf{472}, 33 (2000); Int.J.Mod.Phys. D \textbf{12,} 1015 (2003).

\bibitem{em} E. Elizalde and S. R. Hildebrandt, Phys. Rev.\textbf{\ D} 65
124024 (2004).

\bibitem{maz} P. O. Mazur and E. Mottola. "Gravitational condensate stars:
An alternative to black holes", gr-qc/0109035; "Dark Energy and Condensate
Stars: Casimir Energy in the Large", gr-qc/0405111; Proc. Nat. Acad. Sci.
111 9545 (2004).

\bibitem{br} T. Levi-Civita, Rend. Atti Acad. Naz. Lincei, 2 (1917) 529; I.
Robinson, Bull. Acad. Pol. Sci. 7 (1959) 351; B. Bertotti, Phys. Rev. 116
(1959) 1331.

\bibitem{nar} H. Nariai, Sci. Rep. Tohoku Univ., Ser. 1 \textbf{34}, 160
(1950); \textbf{35}, 62 (1951).

\bibitem{zln} Ya. B. Zel'dovich and I. D.\ Novikov, \textit{Stars and
Relativity} (1971) (University of Chicago Press).

\bibitem{isr} W. Israel, Nuovo Cim. B\textbf{\ 44}, 1 (1966), Errat.-ibid. 
\textbf{48}, 463 (1967).

\bibitem{visgr} M. Visser and D. L. Wiltshire, Class. Quant. Grav. \textbf{21%
}, 1135 (2004), C. Cattoen, T. Faber and M. Visser, Gravastars must have
anisotropic pressures, gr-qc/0505137.

\bibitem{sdust} P. S. Letelier, Phys.\ Rev. D \textbf{20,} 1294 (1979); N.
Dadlich, \textit{On product spacetime with 2-sphere of constant curvature},
gr-qc/0003026.

\bibitem{vf} O. B. Zaslavskii, Phys. Rev. D \textbf{70} 104017 (2004).

\bibitem{rub} V. A. Ruban, JETP Lett. \textbf{8,} 414 (1968) [reprinted:
Gen. Rel. Grav. \textbf{33,} 369 (2001)]; Sov. Phys. JETP \textbf{29,} 1027
(1969) [reprinted in Gen. Rel. Grav. \textbf{33}, 375 (2001)].

\bibitem{nov2} I. D. Novikov, Commun. of the Shternberg State Astronom.
Inst. \textbf{132, }43 (1964).

\bibitem{mt} M. S. Morris and K. S. Thorne, Am. J. Phys. \textbf{56}, 395
(1988).

\bibitem{vis} M.\ S. Visser, \textit{Lorentzian wormholes: From Einstein to
Hawking} (AIP Press, New York, 1995).

\bibitem{phz} O. B. Zaslavskii, D \textbf{72} 061303(R) (2005).

\bibitem{visgen} D. Hochberg and M. Visser, Phys. Rev. D \textbf{56}, 4745
(1997).

\bibitem{dav47} M. Visser and D. Hochberg, Generic wormhole throats,
gr-qc/9710001.

\bibitem{antitr} D. Hochberg and M. Visser, Phys. Rev. D \textbf{58}, 044021
(1998).

\bibitem{back} J. Matyjasek and O. B. Zaslavskii, Phys.\ Rev. D \textbf{71}
087501 (2005).

\bibitem{vit} V. Cardoso, O. J.C. Dias, and J. P.S. Lemos, Phys.Rev. D 
\textbf{70} 024002 (2004).

\bibitem{gl} Guendelman, Gen. Rel. Grav., \textbf{23} (1991) 1415.

\bibitem{flux} V. Dzhunushaliev, U. Kasper and D. Singleton, Phys.Lett. B 
\textbf{479 }(2000) 249; V. Dzhunushaliev, Int.J.Mod.Phys. D14 (2005) 1293.
\end{thebibliography}
\end{document}